\documentclass[preprint,preprintnumbers]{revtex4}

\usepackage{graphicx}

\begin{document}



\title{Dirac's method for constraints: An application to quantum wires}

\author{D.Schmeltzer}
\affiliation{Department of Physics,\\City College of the City University of New York,\\
New York,NY,10031}
\date{\today}

\begin{abstract}

We investigate the  Hubbard model  in the limit $U=\infty$, which is equivalent 
to the  statistical condition of exclusion of double occupancy.
We solve this problem using Dirac's method for constraints.  The constraints are solved within  the Bosonization method. We find that the constraints modify the anomalous commutator.

We apply this theory to quantum wires  at finite temperatures where  the Hubbard  interaction  is $U=\infty$. We  find that the anomalous commutator  induced by the constraints gives rise to the $0.7$   anomalous conductance.



\end{abstract}


\maketitle

\textbf{1. Introduction}

\vspace {0.2 in}

 Advances in the physics of electronic devices  requires the computation of   the wave function  for confined   geometries  and strong interactions. A possible way to  study such cases is the use of quantum   constraints. A particular situation occurs in  quantum wires when  the short range   electron-electron interaction is governed by a    large repulsive  Hubbard interaction.  When     this interaction obeys  $U\rightarrow\infty$,   double occupancy is prohibited.   This problem,  known in the literature  as the Hubbard model  at $U=\infty$  \cite{Kotliar,Wiegmann,Tungler,slaveD, Anderson,Muthukumar,DavidC1}, can be  studied using the method of Dirac's constraints \cite{Dirac}.

Interestingly the physics of quantum wire seems to   depend on the electron-electron interactions \cite{Kane,Haldane,Safi,Maslov,David}. The physics  of electron-electron interactions in one-dimensional metals  is described by  a $Luttinger$ liquid  which replaces the traditional  Fermi liquid description. 
For large  Hubbard interactions, $U\rightarrow\infty$  at finite temperatures, one obtains    an  $incoherent$ $Luttinger$ $liquid$    \cite{Fiete,Cheianov} characterized by    magnetic excitations which are   negligible in comparison to the temperature.

The physics of the Hubbard model  at $U\rightarrow\infty$ might be  relevant for studies of  the $0.7$ conductance anomaly  discovered in quantum wires \cite{Pepper,Peperoni,Cronennwett,Michael}. Experiments suggest that the anomalous conductance is observed at low electronic densities \cite{Picciotto} and   short in wires  \cite{Reilly}.    
Recent Monte Carlo simulations,  performed in the limit  of large Hubbard $U$  and finite temperature, show that the one-dimensional conductance  is anomalous \cite{Siljuasen}.
In ref.\cite{Han},  the one-dimensional itinerant electron model with ferromagnetic coupling has also been investigated. Performing a Monte-Carlo study, the authors in ref.\cite{Han}  have found that the conductance in this case is anomalous  as well . 

From the   work of ref. \cite{Nagaoka}, we know   that a one dimensional model with nearest neighbor hopping  elements  and $U\rightarrow\infty$ has a ferromagnetic ground state  at zero temperature.
The  strongly interacting Hubbard model $U=\infty$ and the  ferromagnetic interactions    are  both  characterized by the exclusion of double occupancy. Therefore, it   seems  that the anomalous conductance  reported by  \cite{Siljuasen} and  \cite{Han}  might have the same origin.

Experimentally, the  anomalous conductance was    observed  in  GaAs/AlGaAs   at low electronic density where   the long range Coulomb interaction plays  a significant role  \cite{Shultz,Matveev,Meyer,Haverim}. In  ref. \cite{Haverim}, it has been shown  that   the long range Coulomb interactions  in the limit of low electronic density drives  the one dimensional wire to the  $U\rightarrow\infty$ limit.

Inspired by   the Monte-Carlo results   \cite{Siljuasen,Han},  we will  compute the conductance for  a one dimensional wire in  the limit   $U=\infty$
at finite temperature using the method of Dirac's constraints.  The   limit $U=\infty$ gives rise  to  exclusion of double occupancy  (this means that in the ground state the fermion occupation number takes only two values , zero or one ). Therefore,  the statistics is similar to  the statistics of  spinless electrons. 
The exclusion of double occupancy  is taken into  consideration,  by demanding that the ground state wave function  $|F>$ must be annihilated by the pair operator $\Psi_{\downarrow}(x)\Psi_{\uparrow}(x)|F>=0$ (  $\Psi_{\sigma}(x)$ is the single electron operator). 
Following Dirac  \cite{Dirac,DavidC1,DavidC2}, we learn that  the constraints must be satisfied at any time. As a result, one   generates a  set of  constraints    which satisfy the equation  $Q_{i}(\vec{x})|F>=0$ ,$i=1,2,3 ..$

In the  language of quantum mechanics  this means that  one has to find   the  wave function $|F>$  which is annihilated by the  set of constraints $Q_{i}(\vec{x})|F>=0$, $i=1,2,3..$ and is an eigenstate of the   Hamiltonian   $H|F>=E|F>$.  When the commutator of the constraints 
$[Q_{i}(\vec{x}),Q_{j}(\vec{x'})]$  with $i\neq j$  can be inverted, the commutation relations are modified to the Dirac commutators \cite{DavidC1,DavidC2}.
As a result, the Heisenberg equation of motion for an  observable $A$ is modified from 
$i\hbar \frac{dA }{dt}=[A,H]$ to $i\hbar \frac{dA }{dt}=[A,H]_{D}$  where $[A,H]_{D}$  is  the Dirac commutator given  by 
$[A,H]_{D}\equiv [A,H]-[A,Q_{1}]([Q_{1},Q_{2}])^{-1}[Q_{2},H]-[A,Q_{2}]([Q_{2},Q_{1}])^{-1}[Q_{1},H]$.
The  presence of the constraints  $[Q_{1},Q_{2}]$ (in the Dirac  commutator)  modifies     the   Pauli statistics  to  the spinless electrons statistics.

For $U=\infty$, the Luttinger liquid at finite  temperature  is replaced by the incoherent Luttinger liquid characterized by  the pinned   spin excitations and  propagating spinless electron.  Therefore,  only the electron current is  conserved.
As a result, when an electron with spin up   or spin down is injected at one end of the wire, we find that the electron which reaches the other end of the wire has an arbitrary  spin.  The interactions are  restricted to the wire and are absent in the leads.  Therefore, 
the density in the wire  corresponds to the spinless electron density which is  half the  density  of the  non-interacting electrons in the leads. This change of  the density will modify  the commutation equation  for the Bosonic densities  by a factor of two. The effect of this modification is observed  at a  finite temperature when a voltage difference  is applied between  the two reservoirs which are connected to the   wire-leads system.   The injected current from the reservoirs will be twice the transmitted current  into the wire.
 The interface between the leads and wire    randomizes the spin and causes the electrons to behave  as spinless  particles at finite temperatures in the wire.
  Therefore, the conductance  will be half in comparison with a non -interacting wire. 

The plan of this paper is as follows:    For pedagogical reasons, we present in chapter $2$ the method of Dirac's constraints used in  quantum mechanics.  This method will be used in the remaining chapters.  In chapter 3,   we present the model  for  an interacting wire ($U=\infty$) of length $d$  coupled  to the non-interacting  leads and reservoir ($U=0$).   In chapter   $4$,  we  derive the  anomalous Dirac commutator for the quantum wire model. This is done  by extending  the   Bosonization method to   the  constraints .  In chapter $5$, we identify the  electric current operator and in chapter $6$, we compute the static current at a finite temperature.    In chapter  $7$, we consider the effect of the Zeeman interaction,
and in chapter  $8$,  we present the conclusion.

\vspace{0.2 in}

\textbf{2. Dirac's method  in quantum mechanics}

\vspace{0.2 in}
  
 In  this section, we will present  Dirac's method in quantum mechanics. This method will be extended  in the next sections to the strongly interacting quantum wires.  
 
  We  consider the   Hamiltonian $H$ with the  constraint  operator $Q_{1}$. The constraint operator is defined by the condition that when it acts on the true ground state $|F>$ it satisfies the equation  $Q_{1}|F>=0$. Formally this  condition is introduced by demanding that 
the ground state wave function  $|F>$  of the Hamiltonian $H$ has to obey, in addition the constraint equation, $Q_{1}|F>=0$ .   Since the constraint must be obeyed at any time, we must have   $\frac{dQ_{1}(t)}{dt}|F>=0$. From the Heisenberg equation of motion,  we obtain the condition  $i\hbar\frac{dQ_{1}(t)}{dt}|F>= [Q_{1}(t),H]|F>=0$. Therefore, we find that the constraint is satisfied at any time, only if the constraint operator $Q_{1}$ commutes with the Hamiltonian  $[Q_{1}(t),H]$ or satisfies the equation     $[Q_{1}(t),H]\propto Q_{1}(t)$. If this is the case,  $Q_{1}$ will be the only constraint.
Here we will assume that this not the case!  Therefore, the only way to satisfy the condition  $\frac{dQ_{1}(t)}{dt}|F>=0$  is  to  introduce  a second constraint $Q_{2}\propto [Q_{1},H]$ (or $Q_{2}\propto [Q_{1},H]-constant\cdot Q_{1}$ ).  For simplicity we assume that no more constraints are needed. As a result,  we obtain a problem with two constraints    $Q_{1}|F>=0$, and   $Q_{2}|F>=0$ which do not commute   $[Q_{1},Q_{2}]\neq 0$.  To enforce the  constraints on the wave function, we use  the method of   Lagrange multipliers  $\lambda_{1}$ and $\lambda_{2}$.\hspace{10 in} 
This allows us to replace  the Hamiltonian  $H$   by
$H_{T}=H+\lambda_{1}Q_{1}+ \lambda_{2}Q_{2}$. \hspace{4 in}
We will determine the Lagrange multipliers  using  the   Heisenberg equations of motion.  \hspace{4 in}
From $i\hbar \frac{d Q_{1}(t)}{dt}|F>=([Q_{1},H]+ \lambda_{2}[Q_{1},Q_{2}])|F>=0$, 
we obtain $ \lambda_{2}=-([Q_{1},Q_{2}])^{-1}[Q_{1},H]$, \hspace{0.1 in} and from
$i\hbar \frac{d Q_{2}(t)}{dt}|F>=([Q_{2},H]+ \lambda_{1}[Q_{1},Q_{2}])|F>=0$, we obtain $ \lambda_{2}=-([Q_{1},Q_{2}])^{-1}[Q_{1},H]$.\hspace{0.1 in}
We substitute  the explicit form of the Lagrange multipliers into the Hamiltonian $H_{T}$, and  find the following Heisenberg equation of motion for $any$ operator $A$:
\begin{equation}
i\hbar \frac{dA }{dt}=[A,H_{T}]= [A,H]-[A,Q_{1}]([Q_{1},Q_{2}])^{-1}[Q_{2},H]-[A,Q_{2}]([Q_{2},Q_{1}])^{-1}[Q_{1},H]\equiv[A,H]_{D}
\label{Diracc}
\end{equation}
The conclusion from this calculation is that the effect of the constraints have changed the regular commutator $[A,H]$ to the Dirac bracket commutator  $[A,H]_{D}$, which is given by  $[A,H]_{D}\equiv [A,H]-[A,Q_{1}]([Q_{1},Q_{2}])^{-1}[Q_{2},H]-[A,Q_{2}]([Q_{2},Q_{1}])^{-1}[Q_{1},H]$. 

Using this result, we define the $Dirac$ $bracket$   for two operators $A$  and $B$ in the following way: 

\begin{equation}
[A,B]_{D}\equiv [A,B]-[A,Q_{1}]([Q_{1},Q_{2}])^{-1}[Q_{2},B]-[A,Q_{2}]([Q_{2},Q_{1}])^{-1}[Q_{1},B]
\label{eqAB}
\end{equation}

This methodology will be applied in the next sections to the problem of strongly interacting electrons.

\vspace{0.2 in}

\textbf{3. The  model for a short wire with exclusion of double occupancy coupled to non-interacting  leads and reservoir }

\vspace{0.2 in}

We  consider a  quantum wire of length $d$ which is perfectly coupled to the  leads of length $L-d$  shown in figure 1.  The leads are  in thermal equilibrium  with the  two electronic reservoirs  which have   chemical potentials  $\mu_{R}$  and  $\mu_{L}$. The reservoirs are described by three dimensional non-interacting electrons. Physically, once the  electrons reach the reservoir they do not return to the wire .  At a finite temperature $T$, the infinite leads  can be replaced by  finite leads  $L\approx L_{T}$ ($L_{T}$ is the thermal  length for which  coherency in the leads is  preserved). 
At  finite temperatures we have the  model: $ H=H_{wire}+H_{leads}+H_{reservoir}$

\begin{eqnarray}
&&H_{wire}+H_{leads}=\int_{\frac{-L}{2}}^{\frac{L}{2}}\,dx[ \sum_{\sigma=\uparrow,\downarrow}
\frac{-\hbar^{2}}{2m}\Psi^{\dagger}_{\sigma}(x)\frac{\partial^{2}}{\partial  x^2}\Psi_{\sigma}(x) +U(x)n_{\sigma=\uparrow}(x)n_{\sigma=\downarrow}(x)-E_{F}\sum_{\sigma=\uparrow,\downarrow}\Psi^{\dagger}_{\sigma}(x)\Psi_{\sigma}(x)]\nonumber\\&&
\end{eqnarray} 

$U(x)$ is the  space dependent Hubbard interaction, $n_{\sigma}(x)=\Psi^{\dagger}_{\sigma}(x)\Psi_{\sigma}(x)$  represents the electronic density  for the spin polarization  $\sigma=\uparrow,\downarrow$ and $E_{F}$ is   the Fermi energy.\hspace{5.0in}
The Hubbard interaction  $U(x)$ is restricted  to   the region   $|x|\leq\frac{d}{2}$ (with the condition $d<<L$).
In the complimentary region $|x|>\frac{d}{2}$   the Hubbard interaction is  $U(x)=0$. 

In the remaining part of this section we will present the  Bosonized version  of two models which have the same Hamiltonian and differ only by the constraint conditions. The two models are: the non interacting electrons described by $U=0$ and the strongly interacting electrons   $U=\infty$. Both problems will be investigated using the method of Bosonization.
The method of Bosonization is based on the representation of  the Fermion operator $\Psi_{\sigma}(x)$ in terms of the  Bosonic fields  $\vartheta_{R,\sigma}=\frac{\vartheta_{\sigma}-\varphi_{\sigma}}{2}$, $\vartheta_{L,\sigma}=\frac{\vartheta_{\sigma}+\varphi_{\sigma}}{2}$.
The Bosonic representation of the electron  operator is given by  \cite{Shankar,Avadh,Mathieu} :

\begin{equation}
\Psi _{\sigma}(x)=\frac{1}{\sqrt{2\pi a}}[e^{iK_{F}x}e^{i\sqrt{4 \pi}\vartheta_{R,\sigma}(x)}+ e^{-iK_{F}x}e^{-i\sqrt{4 \pi}\vartheta_{L,\sigma}(x)}] 
\label{fermion}
\end{equation}
where $a$ is the lattice constant  and $K_{F}$ is the Fermi momentum.
Using  the Bosonic fields   $\vartheta_{L,\sigma}(x)$ and $\vartheta_{R,\sigma}(x)$,  we define a new   field  $P_{\sigma}(x)\equiv\partial_{x}\varphi _{\sigma}(x)$. In order to construct a quantum theory, one has to identify the canonical momentum operator which is conjugate to the Bosonic fields  $\vartheta_{\sigma}(x)$. Here, we have an example  where both the  operator  $\vartheta_{\sigma}(x)$ and   $P_{\sigma}(x)\equiv\partial_{x}\varphi _{\sigma}(x)$ (the candidate for conjugate momentum) are built from the same fields ( $\vartheta_{R,\sigma}$ ,$\vartheta_{L,\sigma}$ ). Therefore, the commutator  $[\vartheta_{\sigma}(x),P_{\sigma'}(x')]$ can be finite    only for an infinite number of particles and  vanishes otherwise. Such commutators are called anomalous \cite{Shankar,Mathieu}. For such cases the commutator 
$[\vartheta_{\sigma}(x),P_{\sigma'}(x')]$ will be defined according to the many particle ground state  $|G>$  with the expectation value  $<G|[\vartheta_{\sigma}(x),P_{\sigma'}(x')]|G>$. 

\vspace{0.1 in}

\textbf{A-The non-interacting electrons, $U(x)=0$}

\vspace{0.1 in} 

The Hamiltonian for the non-interacting electrons is given by eq.3 with $U(x)=0$. The Bosonized Hamiltonian is given by :
\begin{equation}
H^{(U(x)=0)}=\sum_{\sigma=\uparrow,\downarrow}\int_{\frac{-L}{2}}^{\frac{L}{2}}\,dx[\frac{\hbar v}{2}[(\partial_{x}\varphi _{\sigma}(x))^2+(\partial_{x}\vartheta_{\sigma}(x))^2] + \mu_{L}n_{L,\sigma}(x)+ \mu_{R}n_{R,\sigma}(x)]
\label{no}
\end{equation}
where $ n_{R,\sigma}(x)$ and $n_{L,\sigma}(x)$ are the  right and left electronic densities and $\mu_{R}$, $\mu_{L}$ are the chemical potentials.
The non-interacting electrons  ground state is identified with the Fermi sea $|F^{(0)}>$.
 We compute   the  expectation value of the operators   $[\vartheta_{\sigma}(x),P_{\sigma'}(x')]$  and identify   the anomalous commutator   \cite{Shankar,Mathieu}:

\begin{equation}
 <F^{(0)}|[\vartheta_{\sigma}(x),P_{\sigma'}(x')]|F^{(0)}> =i\delta_{\sigma,\sigma'}\delta(x-x')
\label{com}
\end{equation}

The result in equation $(6)$  allows to identify the operator $P _{\sigma}(x)\equiv\partial_{x}\varphi _{\sigma}(x)$ as the   canonical conjugated momenta operator  and define   the  commutator $[\vartheta_{\sigma}(x),P_{\sigma'}(x')]= i\delta_{\sigma,\sigma'}\delta(x-x')$.
(Once the commutators have been defined we can choose the  representation       representation:  $\vartheta_{\sigma}(x)|\vartheta_{\uparrow},\vartheta_{\downarrow}>=\vartheta_{\sigma}(x)|\vartheta_{\uparrow},\vartheta_{\downarrow}>$  with  the conjugated momenta  given  by 
$P _{\sigma}(x)=-i\frac{\delta}{\delta \vartheta_{\sigma}(x)}$.)

\vspace{0.1 in}

\textbf{B-The strongly interacting electrons  $U(x)=\infty$ }

\vspace{0.1 in}

We consider the strongly interacting case described by the Hubbard interaction $U(x)=\infty$.  ( For finite Hubbard $U$ interactions, the  Coulomb  interactions   \cite{Malard}   drive  the the Hubbard $U$ interaction to $U\rightarrow\infty$. The limit $U(x)=\infty$  describes effectively the physics in the strong coupling limit )

The Hubbard interaction  $U(x)=\infty$ is  restricted 
to  the region   $|x|\leq\frac{d}{2}$. As a result the  the electron  occupation number  is restricted  to the charge values    $n_{e}(x)=0,1$.   

\begin{equation}
\Psi^{\dagger}_{\sigma=\uparrow}(x)\Psi_{\sigma=\uparrow}(x)+ \Psi^{\dagger}_{\sigma=\downarrow}(x)\Psi_{\sigma=\downarrow}(x)|F>=n_{e}(x)|F>
\label{cp}
\end{equation}
Where   $|F>$ is the electronic ground state which is different from the non-interacting ground state $|F^{(0)}>$.
This condition is named  the exclusion of double occupancy and is implemented  with the help of the  constraint equation.

\begin{equation}
\Psi_{\downarrow}(x)\Psi_{\uparrow}(x)|F>=0   \hspace{0.1 in} for \hspace{0.05 in} the \hspace{0.05 in} region  \hspace{0.1 in} |x|\leq \frac{d}{2}
\label{frac}
\end{equation} 

In the complimentary region 
$|x|>\frac{d}{2}$  the electrons are non-interacting,   and we have $U(x)=0$.

 Once the constraint conditions have  been introduced we can use the   non-interacting Hamiltonian  for the wire-leads  system.

\begin{equation}
H_{wire}+H_{leads} \approx\int_{\frac{-L}{2}}^{\frac{L}{2}}\,dx[ \sum_{\sigma=\uparrow,\downarrow}
\frac{-\hbar^{2}}{2m}\Psi^{\dagger}_{\sigma}(x)\frac{\partial^{2}}{\partial  x^2}\Psi_{\sigma}(x) -E_{F}\sum_{\sigma=\uparrow,\downarrow}\Psi^{\dagger}_{\sigma}(x)\Psi_{\sigma}(x)]
\label{eq}
\end{equation}

Next, we Bosonize   this model using the same fields as we used for the noninteracting case.
In spite of the fact that the Hamiltonian for the non-interacting case is identical to the strongly interacting one, the two ground states are   different. The new ground state  $|F>$
obeys the constraint equation $(8)$ which, in the Bosonic form, is given by the representation :



\begin{equation}   
\Psi_{\uparrow}(x) \Psi_{\downarrow}(x)|F>= 2e^{-i\sqrt{4\pi}\varphi_{e}(x)}cos[2K_{F}x +\sqrt{\pi}\vartheta_{e}(x)]+cos[\sqrt{2\pi}\vartheta_{s}(x)]|F>=0; \hspace{0.1 in}   |x|\leq\frac{d}{2} 
\label{constraints}
\end{equation}  
The constraint operator is represented in terms of the Bosonic fields  for the charge degrees of freedom:  $\vartheta_{e}(x)=(\vartheta_{\uparrow}(x)+\vartheta_{\downarrow}(x))$ ,  $\varphi _{e}(x)=\frac{1}{2}(\varphi _{\uparrow}(x) +\varphi _{\downarrow}(x))$  and   $P_{e}(x)=\partial_{x}\varphi _{e}(x)$. Similarly, the spin degrees of freedom are given by: $\vartheta_{s}(x)=\frac{1}{\sqrt{2}}(\vartheta_{\uparrow}(x)-\vartheta_{\downarrow}(x))$,  $\varphi _{s}(x)=\frac{1}{\sqrt{2}}(\varphi _{\uparrow}(x) -\varphi _{\downarrow}(x))$  and  $P_{s}(x)=\partial_{x}\varphi _{s}(x)$.

In order to find the commutation relation, we will use the new ground state which is restricted by the constraint. The commutator  $[\vartheta_{\sigma}(x),P_{\sigma'}(x')]$  will be defined by the value of the expectation value with respect the ground  state  \cite{Shankar,Mathieu}. This expectation value is  the central result of this paper and is derived in the next section.  The expectation value of the commutator is a function of  $\textbf{h}(x,d)$ given by:
$\textbf{h}(x,d)=1$ for   $|x|\leq\frac{d}{2}$,  ($U=\infty$) and $\textbf{h}(x,d)=0$ for $|x|>\frac{d}{2}$,    ($U=0$).

\begin{equation}
 <F| [\vartheta_{e}(x),P _{e}(x')]|F> =i\delta(x-x')[1-\textbf{h}(x,d)\frac{1}{2}] 
\label{com}
\end{equation}
This result allows  us to define the Dirac  commutator $ [\vartheta_{e}(x),P _{e}(x')]_{Dirac}= i\delta(x-x')[1-\textbf{h}(x,d)\frac{1}{2}] $.

The  Bosonized       Hamiltonian  for this case is given by :       
\begin{equation}
H_{wire}=\sum_{\sigma=\uparrow,\downarrow}\int_{\frac{-d}{2}}^{\frac{d}{2}}\,dx\frac{\hbar v}{2}[(\partial_{x}\varphi _{\sigma}(x))^2+(\partial_{x}\vartheta_{\sigma}(x))^2]
\label{bos}
\end{equation}

\begin{equation} H_{Leads}=\sum_{\sigma=\uparrow,\downarrow}[\int_{\frac{-L}{2}}^{\frac{-d}{2}}\,dx\frac{\hbar v}{2}[(\partial_{x}\varphi _{\sigma}(x))^2+(\partial_{x}\vartheta_{\sigma}(x))^2]+ \int_{\frac{d}{2}}^{\frac{L}{2}}\,dx\frac{\hbar v}{2}[(\partial_{x}\varphi _{\sigma}(x))^2+(\partial_{x}\vartheta_{\sigma}(x))^2]
\label{leads}
\end{equation}   
$v$  is the Fermi velocity for the non-interacting fermions.  
We assume  perfect coupling  between the wire and the leads. The Bosonic fields obey continuous   boundary conditions:  $\vartheta_{\sigma}(x=\frac{d}{2}-\epsilon)= \vartheta_{\sigma}(x=\frac{d}{2}+\epsilon)$, $\vartheta_{\sigma}(x=-\frac{d}{2}-\epsilon)
=\vartheta_{\sigma}(x=-\frac{d}{2}+\epsilon)$ and $\varphi_{\sigma}(x=\frac{d}{2}-\epsilon)= \varphi_{\sigma}(x=\frac{d}{2}+\epsilon)$, $\varphi_{\sigma}(x=-\frac{d}{2}-\epsilon)
=\varphi_{\sigma}(x=-\frac{d}{2}+\epsilon)$ , where $\epsilon$ represents the overlapping wire-leads region.\hspace{10 in}
Therefore, we can replace  the Hamiltonian  $H_{wire}+H_{leads}$ by a free Bosonic model.
\begin{eqnarray} &&H_{wire}+H_{leads}=\sum_{\sigma=\uparrow,\downarrow}\int_{\frac{-L}{2}}^{\frac{L}{2}}\,dx\frac{\hbar v}{2}[(\partial_{x}\varphi _{\sigma}(x))^2+(\partial_{x}\vartheta_{\sigma}(x))^2]\nonumber\\&&
H_{reservoir}= \sum_{\alpha=\uparrow,\downarrow}\int_{\frac{-L}{2}}^{\frac{L}{2}}\,dx [\mu_{L}(x)n_{L,\sigma}(x)+ \mu_{R}(x)n_{R,\sigma}(x)]
\end{eqnarray}

$H_{reservoir}$ is the reservoir Hamiltonian which is a function of   the  voltage difference  $\frac{\mu_{R}(x)-\mu_{L}(x)}{-e}=V$  and  the    right and left  electronic densities  
$n_{R,\sigma}(x)\equiv \rho_{R,\sigma}(x)+<F|n_{R,\sigma}(x)|F>$, $n_{L,\sigma}(x)\equiv \rho_{L,\sigma}(x)+<F|n_{L,\sigma}(x)|F>$.  The Bosonic representation of the densities is: $\rho_{R,\sigma}(x)=\frac{1}{2\sqrt{\pi}}(\partial_{x}\vartheta_{\sigma}(x)- \partial_{x}\varphi _{\sigma}(x))$ and     $\rho_{L,\sigma}(x)=\frac{1}{2\sqrt{\pi}}(\partial_{x}\vartheta_{\sigma}(x)+ \partial_{x}\varphi _{\sigma}(x))$ \cite{Avadh,Mathieu}. Due to the fact that the commutator for the electronic densities  is space  dependent,  the average   density is space dependent  $<F|n_{R,\sigma}(x)+n_{L,\sigma}(x)|F>=n_{e}[1-\frac{\textbf{h(x,d)}}{2}]$  where     $n_{e}=\frac{1-\delta}{a}$ is the electronic density in the leads. 
The space dependent electrostatic potential caused by the space dependent charge density  is given by  $\delta v(x)=\frac{\mu_{R}(x)+\mu_{L}(x)}{(-e)}$  \cite{datta}.
This potential will introduce current fluctuations.

\vspace{0.2 in}

\textbf{4. Dirac's method for the exclusion of double occupancy-an application to the finite wire-leads  system-A Bosonization formulation for  the constraints}

\vspace{0.2 in}

 We will extend  the  Quantum Mechanical results obtained in the previous section to the wire Hamiltonian   $H_{wire}$ and we will introduce  a  new formulation for the constraints using the method of Bosonization. 
For the  wire Hamiltonian $H_{wire}$, we will enforce the  exclusion of double occupancy. The operator for  exclusion of double occupancy \cite{DavidC1} leads to the constraint condition for the ground state $|F>$ which replaces the non-interacting Fermi sea $|F^{(0)}>$. The exclusion of double occupancy is enforced  by  the two electrons operator   $\Psi_{\uparrow}(x) \Psi_{\downarrow}(x)|F>=0$ for $|x|\leq\frac{d}{2}$ . Except for  half filling, $\Psi_{\uparrow}(x) \Psi_{\downarrow}(x)$ is the only $primary$ constraint \cite{Dirac}. In order to find all the  $secondary$ constraints \cite{Dirac}, we have to commute the constraint operator  $\Psi_{\uparrow}(x) \Psi_{\downarrow}(x)$ with the Hamiltonian. If this commutator is not equal to the constraint field,  new constraints are generated.
Using the Bosonic representation  $\Psi_{\sigma}(x)$  given in equation $(5)$, we compute the representation of the pair operator  $\Psi_{\uparrow}(x)\Psi_{\downarrow}(x)$.
The pair operator is represented   in terms of the Bosonic fields,  $\vartheta_{e}(x)$  ,  $\varphi _{e}(x)$,$\vartheta_{s}(x)$ and $\varphi _{s}(x)$:
\begin{equation}   
\Psi_{\uparrow}(x) \Psi_{\downarrow}(x)|F>= 2e^{-i\sqrt{4\pi}\varphi_{e}(x)}Q_{1}(x)|F>=0; \hspace{0.1 in}   |x|\leq\frac{d}{2} 
\label{constraints}
\end{equation}   
 where   $Q_{1}(x)$ is  the Bosonic   constraint  : 
\begin{equation}
Q_{1}(x)=cos[2K_{F}x +\sqrt{\pi}\vartheta_{e}(x)]+cos[\sqrt{2\pi}\vartheta_{s}(x)];\hspace{0.2 in} Q_{1}(x)|F>=0
\label{coperator}
\end{equation}
In order to understand the effect of the constraints, we will present a simplified description.  For this purpose we will ignore the $2K_{F}$ oscillations, and  we will  approximate  $Q_{1}(x)$ by  $Q_{1}(x)\approx cos[\sqrt{2\pi}\vartheta_{s}(x)]$.  From the equation $cos[\sqrt{2\pi}\vartheta_{s}(x)|F>\approx0$,
we learn that the spinon  phase $ \vartheta_{s}(x)$ must be pinned to a constant value $\vartheta_{s}(x)= (\sqrt{2\pi})^{-1}[\frac{\pi}{2}+n\pi]$, $n=0,1,2..$. Therefore, the system has no spinon excitations, and the only excitations are spinless fermions.
It is this constraint which gives strong exponential decays of the spinon correlation function. As a result, the system is described effectively by a spinless system. Therefore, due to the exclusion of double occupancy,   the electronic density in the wire will be half in comparison to the leads, where the constraint is absent. This changes   will modify  the commutation equation  for the electron number  from  $[\vartheta_{e}(x),P_{e}(x')]=i\delta(x-x')$ to  a new commutator $[\vartheta_{e}(x),P_{e}(x')]_{D}=\frac{i}{2}\delta(x-x')$.

For the remaining part, we will work with the full constraint operator given in equation $(16)$. The constraint  $Q_{1}(x,t)$ must be satisfied at any  time. Therefore, we must have $\frac{dQ_{1}(x,t)}{dt}|F>=0$.
Using the Heisenberg equations of motion with the non-interacting anomalous commutator given in equation $(6)$, we compute : 
$i\hbar\frac{dQ_{1}(x,t)}{dt}=[H_{wire},Q_{1}(x,t)]$.   The condition  $\frac{dQ_{1}(x,t)}{dt}|F>=0$ can be satisfied only  if one introduces a new constraint  $Q_{2}(x)$  given by  $[Q_{1}(x,t),H_{wire}]$ : 
\begin{equation}
Q_{2}(x)=\sqrt{2}sin[2K_{F}x +\sqrt{\pi}\vartheta_{e}(x)]P_{e}(x)+sin[\sqrt{2\pi}\vartheta_{s}(x)]P_{s}(x)
\label{newconstraint}
\end{equation}
The constraint    must annihilate the ground state, $Q_{2}(x)|F>=0$, and we must also have $\frac{dQ_{2}(x,t)}{dt}|F>=0$.  In order for this to happen, we need to include a third     constraint  $Q_{3}(x)\propto[[Q_{1}(x,t),H_{wire}],H_{wire}]$ which is given by: 
$Q_{3}(x)=sin[2K_{F}x +\sqrt{\pi}\vartheta_{e}(x)] (\sqrt{2}P_{e}(x))^2+sin[\sqrt{2\pi}\vartheta_{s}(x)] P^2_{s}(x)$. Continuing this process, we identify additional constraints \hspace{0.1 in} $Q_{4}(x)$, $Q_{5}(x)$.....  The third order constraint     $Q_{3}(x,t)\propto[[Q_{1}(x,t),H_{wire}],H_{wire}]$  can be written  as a second  order time  derivative  $Q_{3}(x,t)\propto \frac{d^{2}Q_{1}(x,t)}{dt^{2}}$.   The higher order  constraints  $Q_{i}(x,t)$, $i=3,4,5,..$ can be neglected since they are represented   by higher order  time derivatives :  $Q_{i}(x,t)\propto \frac{d^{i-1}Q_{1}(x,t)}{dt^{i-1}}$,  $i\geq3$ \cite{Goldenfeld}.

Following \cite{Dirac,DavidC2}, we find that the operators  $Q_{i}(x)$ i=1,2 form a set of $Second$ $Class$ constraints, $[Q_{1}(x),Q_{2}(x)]\approx <F| [Q_{1}(x),Q_{2}(x)]|F>\neq 0$.
In order to find the new commutation rules for our system, we will use the Lagrange multiplier fields $\lambda_{i}(x)$ 
which will be multiplied by   the step function  $\textbf{h}(x,d)$  (which is one for  $|x|\leq \frac{d}{2}$ and zero otherwise).
We are interested to find the commutation rules for the full wire-leads system. In order to do this,    
we replace the  Hamiltonian $H_{wire}+H_{leads}$ by the the Hamiltonian $H_{T}$ which is a function of the $Lagrange$ fields $\lambda_{i}(x)$  that enforce the  constraints $Q_{i}(x)$   for the wire \cite{Dirac}.  Since we have no constraints for the leads, we will introduce the step  function $\textbf{h}(x,d)$  which  is zero for the leads  and  takes the value of one for the wire.  The  constraints  for  the wire will be enforced by replacing the $Lagrange$ fields $\lambda_{i}(x)$ by the product   $\textbf{h}(x,d)\lambda_{i}(x)$.
\begin{equation}
H_{T}=H_{wire}+H_{leads}+\int_{\frac{-L}{2}}^{\frac{L}{2}}\,dx \textbf{h}(x,d)\sum_{i=1}^{2}\lambda_{i}(x)Q_{i}(x))
\label{heavy}
\end{equation} 
We will apply the quantum mechanical method  given in equation $(9)$   to the  Hamiltonian for the wire given in equation $(18)$. Since the constraint must obey $\frac{d Q_{1}(x,t)}{dt}|F>=0$,  and   the commutator   $[Q_{1}(x),Q_{2}(x')]$ can be inverted,  the $Lagrange$ multipliers can be computed. 
\begin{eqnarray} 
&&i\hbar\frac{d Q_{1}(x,t)}{dt}|F>=[Q_{1}(x,t),H_{wire}+H_{leads}+\int_{\frac{-L}{2}}^{\frac{L}{2}}\,dx\textbf{h}(x,d)\sum_{i=1}^{2}\lambda_{i}(x)Q_{i}(x))]|F>=0\nonumber\\&&
\end{eqnarray}
From equation $(19)$, we obtain the $Lagrange$ fields $\lambda_{i}(x)$. 
The $Lagrange$ fields are  functions of  the matrix     $C_{i,r}(x,u)$, $r=1,2$  which are defined with the help of the the integral equation
$\sum_{r=1}^{2}\int\,du C_{i,r}(x,u)[Q_{r}(u),Q_{j}(y)]=\delta_{i,j}\delta(x-y)$ ,$j=1,2$.
The matrix   $C_{i,r}(x,u)$  is  given by the inverse of the   commutator   $[Q_{i}(x),Q_{r}(x')]$. The commutator  of the constraints    $[Q_{1}(x),Q_{2}(x')]$  (given in equations $(16-17)$) is computed  using the non-interacting  Bosonic commutator  $[\vartheta_{\sigma}(x),P_{\sigma'}(x')]=i\delta_{\sigma,\sigma'}\delta(x-x')$. 
\begin{equation}
[Q_{1}(x),Q_{2}(x')]=-i\sqrt{2\pi}\delta(x-x')[(sin(\sqrt{2\pi}\vartheta_{s}(s)))^2+
(sin(2K_{F}x +\sqrt{\pi}\vartheta_{e}(x)))^2]
\label{comm}
\end{equation}
We substitute in equation $(19)$ the result of the commutators  $[Q_{1}(x),Q_{2}(x')]$ given in  equation $(20)$. As a result,  the equation   $\frac{d Q_{i}(x,t)}{dt}$, $i=1,2$ ( see  equation $(19)$ ) allows  us to determine the   $Lagrange$ fields   $\lambda_{i}(x)$, $i=1,2$.  
Next, we   substitute the $Lagrange$ fields  $\lambda_{i}(x)$ into the total Hamiltonian  $H_{T}$ given in equation  $(18)$ and  find that the Heisenberg equations of motion  are  determined by the physical Hamiltonian $ H_{wire}+H_{leads}$ with  the modified    commutation relation  $[,]_{Dirac}$. 
The Heisenberg equation of motion for any operator $A(x,t)$ will be   given by :  
\begin{eqnarray}
&&i\hbar\frac{d A(x,t)}{dt}=[ A(x,t),H_{wire}+H_{leads}]_{Dirac} \equiv\nonumber\\&&[ A(x,t),H_{wire}+H_{leads}]-\sum_{i=1,2}\sum_{j=1,2}\int\,du \textbf{h}(u,d) \int\,dv  [A(x,t),Q_{i}(u)] C_{i,j}(u,v)[Q_{j}(v),H_{wire}+H_{leads}]\nonumber\\&&
\end{eqnarray}   
The Dirac commutator  for the canonical conjugates fields  $\vartheta_{\sigma}(x)$  and  $P _{\sigma'}(y)$ will be    given by: 
\begin{eqnarray}
[\vartheta_{\sigma}(x), P _{\sigma'}(y)]_{Dirac}&=&[\vartheta_{\sigma}(x), P _{\sigma'}(y)]-\int_{\frac{-L}{2}}^{\frac{L}{2}}\,du \textbf{h}(u,d) \int_{\frac{-L}{2}}^{\frac{L}{2}}\,dv[\vartheta_{\sigma}(x),Q_{2}(u)]C_{2,1}(u,v)[Q_{1}(v), P _{\sigma'}(y)]\nonumber\\&&
\end{eqnarray}
We observe that the Dirac commutator given in equation $(21)$ has exactly the same structure as  equation   $(2)$  obtained in chapter  2.
 
The explicit  form of the  Dirac  commutator  for the electronic density  operator  $\vartheta_{e}(x)$ and  the conjugate field  $P _{e}(x')$,
$[\vartheta_{e}(x),P _{e}(x')]_{Dirac}$ will be  given by:
\begin{eqnarray}
&&[\vartheta_{e}(x),P _{e}(x')]_{Dirac}=
[\vartheta_{e}(x),P _{e}(x')](1-\frac{\textbf{h}(x,d)(sin(2K_{F}x +\sqrt{\pi}\vartheta_{e}(x)))^{2}}{(sin(\sqrt{2\pi}\vartheta_{s}(x)))^{2}+(sin(2K_{F}x +\sqrt{\pi}\vartheta_{e}(x)))^{2}})\nonumber\\&&
\end{eqnarray}

The constraint operator $Q_{1}(x)|F>=0$ causes  any operator $R(x)$  to obey the  equation:  $R(x)Q_{1}(x)|F>=0$. We choose  $R(x)= cos[2K_{F}x +\sqrt{\pi}\vartheta_{e}(x)]-cos[\sqrt{2\pi}\vartheta_{s}(x)]$ and construct the  product  $R(x)Q_{1}(x)=(cos[2K_{F}x +\sqrt{\pi}\vartheta_{e}(x)])^2-(cos[\sqrt{2\pi}\vartheta_{s}(x)])^2$.  Adding and subtracting one from $ R(x)Q_{1}(x)$ gives us a  new equation $R_{2}(x)$  which obeys:  
\begin{equation}
R_{2}(x)|F>\equiv[(sin(\sqrt{2\pi}\vartheta_{s}(x)))^{2}-(sin(2K_{F}x +\sqrt{\pi}\vartheta_{e}(x)))^{2}]|F>=0
\label{equation}
\end{equation} 
This gives as the result: 
\begin{eqnarray}
&&\frac{(sin(2K_{F}x +\sqrt{\pi}\vartheta_{e}(x)))^{2}}{(sin(\sqrt{2\pi}\vartheta_{s}(x)))^{2}+(sin(2K_{F}x +\sqrt{\pi}\vartheta_{e}(x)))^{2}})|F>=\nonumber\\&&\frac{(sin(2K_{F}x +\sqrt{\pi}\vartheta_{e}(x)))^{2}}{2(sin(2K_{F}x +\sqrt{\pi}\vartheta_{e}(x)))^{2} [1+\frac{((sin(\sqrt{2\pi}\vartheta_{s}(x)))^{2}-(sin(2K_{F}x +\sqrt{\pi}\vartheta_{e}(x)))^{2}}{(sin(2K_{F}x +\sqrt{\pi}\vartheta_{e}(x)))^{2}}]}|F> =\frac{1}{2}\nonumber\\&&
\end{eqnarray}
The result in equation $(25)$ was obtained after expanding the last expression in powers of 
  $(\frac{R_{2}(x)}{(sin(2K_{F}x +\sqrt{\pi}\vartheta_{e}(x)))^{2}}])^{n}$  and using  the condition  $(R_{2}(x))^{n}|F>=0$  
The  commutator  $ [\vartheta_{e}(x),P _{e}(x')]$ is finite    only for an infinite number of particles and  vanishes otherwise.  The anomalous commutator will be given    by the expectation value $<F| [\vartheta_{e}(x),P _{e}(x')]|F>$ \cite{Shankar,Mathieu}. This  allows  us to define the \textbf{Dirac commutator} $ [\vartheta_{e}(x),P _{e}(x')]_{Dirac}$ .

\begin{eqnarray}
&&[\vartheta_{e}(x),P _{e}(x')]_{Dirac}\equiv<F|[\vartheta_{e}(x),P _{e}(x')]_{Dirac}|F>\nonumber\\&&=i
\delta(x-x')[1-\textbf{h}(x,d)<F|\frac{(sin(2K_{F}x +\sqrt{\pi}\vartheta_{e}(x)))^{2}}{(sin(\sqrt{2\pi}\vartheta_{s}(x)))^{2}+(sin(2K_{F}x +\sqrt{\pi}\vartheta_{e}(x)))^{2}})|F>]\nonumber\\&&=i
\delta(x-x')[1-\frac{1}{2}\textbf{h}(x,d)]
\end{eqnarray} 
This result shows that the  anomalous commutator (Dirac commutator)  for the wire has been modified to $\frac{ i}{2}\hbar$. This is consistent with fact that the electronic density for  the  wire is  half the density  of the leads (see figure 1).

The  Dirac commutator affect in a significant way the equation of motion for the particle -hole excitations.  Using   the Heisenberg equations of motion $i\hbar\frac{d \vartheta_{e}(x,t)}{dt}=[\vartheta_{e}(x,t),H_{wire}]_{Dirac}$ and $i\hbar\frac{d \varphi_{e}(x,t)}{dt}=[\varphi_{e}(x,t),H_{wire}]_{Dirac}$  we find:

\begin{equation}
[\partial^{2}_{t}-v^{2}(1-\frac{\textbf{h}(x,d)}{2})^{2}\partial^{2}_{x}]\vartheta_{e}(x,t)=0
\label{scat}
\end{equation}
The space dependent commutation rules give rise to a multi particle-hole  scattering state. 
Performing a space average, we obtain the effective  velocity  $v_{eff.}\equiv  v \sqrt{<(1-\frac{\textbf{h}(x,d)}{2}))^2>_{space-average}}=v(1-\frac{3d}{4L})$  with the fundamental   frequency $\omega=v_{eff.}\frac{2\pi}{L}$,. The space dependent part in equation  $(27)$  gives rise to  the scattering potential $V_{sc}(x)\equiv v^2[<(1-\frac{\textbf{h}(x,d)}{2}))^2>_{space-average}-(1-\frac{\textbf{h}(x,d)}{2}))^2]$ which generates high harmonic states.

To conclude the anomalous commutator for a wire with exclusion of double occupancy given in equation $(26)$ is the central result of this paper. This result was obtained restricting the infinite number of  generated constraints  to a finite set of constraints $Q_{1}(x)$ and $Q_{2}(x)$.  We have shown that according to scaling theory  \cite{Goldenfeld}, the higher order constraints  are irrelevant  and therefore can be neglected.


\vspace{0.2 in}

\textbf{5. The current operator for the wire} 

\vspace{0.2 in}

Using   the Dirac commutator,  we   find the  equation of motion for the  electronic density $\rho_{e}(x)\equiv\frac{1}{\sqrt{\pi}}\partial_{x}\vartheta_{e}(x)$.
\begin{eqnarray}
&&\frac{d \rho_{e}(x,t)}{dt}=\frac{1}{i\hbar}[\rho_{e}(x,t),H_{wire}+H_{leads}]_{Dirac}\nonumber\\&&=
\frac{\hbar v}{i\hbar \sqrt{\pi}}\int_{\frac{-L}{2}}^{\frac{L}{2}}\,dy[\partial_{x}\vartheta_{e}(x), P^{2}_{e}(y,t) _{e}+\frac{1}{4}(\partial_{y}\vartheta_{e}(y,t))^2 +\frac{1}{2}( P^{2}_{s}(y,t) +(\partial_{y}\vartheta_{s}(y,t))^2)]_{Dirac}=\nonumber\\&&-\frac{v}{\sqrt{\pi}}\partial_{x}( \int_{\frac{-L}{2}}^{\frac{L}{2}}\,dy[\vartheta_{e}(x),P _{e}(y,t)]_{Dirac}P _{e}(y,t))  \equiv -\partial_{x}J_{e}(x,t)
\end{eqnarray}
We identify from   the continuity equation $\frac{d \rho_{e}(x,t)}{dt}+\partial_{x}J_{e}(x,t)=0$      the conserved  electronic  current operator  $J_{e}(x)$. We   multiply $J_{e}(x)$ by the electric  charge $(-e)$ and find the electric current operator  $\hat{I}_{e}$.
\begin{equation}
\hat{I}_{e}=\frac{ev}{\sqrt{\pi}}\int_{\frac{-L}{2}}^{\frac{L}{2}}\,dy[\vartheta_{e}(x),P _{e}(y)]_{Dirac} P _{e}(y) = \frac{ev}{\sqrt{\pi}} [1-\frac{\textbf{h}(x,d)}{2}]\partial_{x}\varphi_{e}(x)
\label{current}
\end{equation}
This equation shows  that the current in the wire is density dependent.  We find that the current in the wire  ($|x|\leq \frac{d}{2}$)  is given by        $\hat{I}_{e}=\frac{ev}{2\sqrt{\pi}}\partial_{x}\varphi_{e}(x)$
 is reduced by a factor of two in comparison with the non-interacting electrons  ($|x|>\frac{d}{2}$) $ \hat{I}^{(U=0)}_{e}= \frac{ev}{\sqrt{\pi}}\partial_{x}\varphi_{e}(x)$.

Contrary to the non-interacting wire where  both the  charge and spin currents are conserved, the constraints in the wire  allow only for  the conservation of the  electric current $\hat{I}_{e}$. 
Due to the fact that the  commutation equations and the  conservation laws for the currents in the leads and in  the wire are not the same, we have at the interface  between the wire and  the leads incoherent scattering. (Electrons  with a well defined spin (spin up or spin down),  are  injected from  the reservoirs. The  leads-wire interface   randomizes the spin and causes the electrons to behave  as spinless  particles at finite temperatures.)

\vspace{0.2 in}

\textbf{6. The current at finite temperatures for a finite wire $L>>d$ and $L>L_{T}$}

\vspace{0.2 in}

In this section we consider explicitly the limit of $U\rightarrow\infty$ taken at finite temperature $T$. In this limit,   spin exchange  processes are  suppressed  and the only  allowed processes  are the particles-holes (non-zero modes ) and particles excitations  (zero modes). Due to the fact that the system is finite, we can separate the Hamiltonian, and the Bosonic fields into the periodic  non-zero modes Bosons   $\vartheta ^{(n\neq0)}_{R,\sigma}(x)\equiv\hat{\vartheta}_{R,\sigma}(x)$, $\vartheta ^{(n\neq0)}_{L,\sigma}(x)\equiv\hat{\vartheta}_{L,\sigma}(x)$  and the zero mode Bosons  $\vartheta ^{(n=0)}_{R,\sigma}(x)\equiv \frac{x\sqrt{\pi}}{L}N_{R,\sigma}$,  $\vartheta ^{(n=0)}_{L,\sigma}(x)\equiv \frac{x \sqrt{\pi}}{L}N_{L,\sigma}$ where $N_{R,\sigma}$  $N_{L,\sigma}$    measure the added  charges  with respect the Fermi energy  \cite{Avadh,Mathieu}.
  
Using  the  Boson decomposition  we find the  representation:
\begin{equation}
\partial_{x}\varphi_{e}(x)\equiv  \partial_{x}\varphi^{(n\neq0)}_{e}(x)+\partial_{x}\varphi^{(n=0)}_{e}(x)=\partial_{x}\hat{\varphi}_{e}(x)+  \frac{\sqrt{\pi}}{2  L}[N_{\uparrow,L}+N_{\downarrow,L}-N_{\uparrow,R}-N_{\downarrow,R}]
\label{nonzero}
\end{equation}
Similarly we  decompose the  Hamiltonian into  two parts: 
   $H^{(n\neq0)}_{wire}+H^{(n\neq0)}_{leads}=\sum_{\sigma=\uparrow,\downarrow}\int_{\frac{-L}{2}}^{\frac{L}{2}}\,dx\frac{\hbar v}{2}[(\partial_{x}\hat{\varphi} _{\sigma}(x))^2+(\partial_{x}\hat{\vartheta}_{\sigma}(x))^2]$ 
   
and the zero mode part $H^{(n=0)}_{wire}+H^{(n=0)}_{leads}$: 
\begin{equation}
H^{(n=0)}_{wire}+H^{(n=0)}_{leads}=\frac{h v }{2L}\sum_{\sigma=\uparrow,\downarrow}((\frac{d}{L})[N^2_{\sigma,L}+N^2_{\sigma,R}]+\frac{(L-d)}{L}[N^2_{\sigma,L}+N^2_{\sigma,R}])
\label{zero}
\end{equation}
For a constant   external   voltage,  $V=\frac{\mu_{R}(x)- \mu_{L}(x)}{-e}$  we have only the  zero mode  reservoir Hamiltonian:
\begin{equation}
H^{(n=0)}_{reservoir}= \sum_{\sigma=\uparrow,\downarrow}[\mu_{L}N_{\sigma,L}+\mu_{R}N_{\sigma,R}]=\sum_{\sigma=\uparrow,\downarrow}[(\frac{-e V}{2})N_{\sigma,R}- (\frac{-e V}{2})N_{\sigma,L}]
\label{reservoir}
\end{equation}
Since the commutator is space dependent, we will introduce   a  space dependent  electrostatic potential $\delta v (x)=\frac{\mu_{R}(x)+ \mu_{L}(x)}{-e}$ which will couple  to the non-zero mode densities.
The effect of the   electrostatic potential $\delta v (x)$  is introduced using  the    non-zero mode reservoir :
\begin{equation}
H^{(n\neq 0)}_{reservoir}=\int_{-\frac{L}{2}}^{\frac{L}{2}}\,dx[(-e)\delta v (x,t)(n_{e}(x,t)-n_{e})] 
\label{non-reservoir}
\end{equation}
The electronic density is defined with respect the density in the leads: $n_{e}(x,t)-n_{e}\equiv\frac{1}{\sqrt{\pi}}\partial_{x}\vartheta_{e}(x,t)+n_{e}[1-\frac{\textbf{h(x,d)}}{2}]-n_{e}= \frac{1}{\sqrt{\pi}}\partial_{x}\vartheta_{e}(x,t)-n_{e}\frac{\textbf{h(x,d)}}{2}$. 
The electrostatic  potential $\delta v (x,t)$ is  determined by the Poisson equation  \cite{datta}:
\begin{equation}
\delta v (x,t)=\frac{-e}{ \kappa}\int_{-\frac{L}{2}}^{\frac{L}{2}}\,dy[\frac{1}{\sqrt{(x-y)^2+a^2 }}-\frac{1}{\sqrt{(x-y)^2+\xi^2} }][\frac{1}{\sqrt{\pi}}\partial_{y}\vartheta_{e}(y,t)-n_{e}\frac{\textbf{h(y,d)}}{2}]
\label{pot}
\end{equation}
$\kappa$ is the dielectric constant and $\xi$ is the gate screening length \cite{Meyer}. 

Next we will compute the two components  currents $\partial_{x}\varphi^{(n\neq0)}_{e}(x)$  and $\partial_{x}\varphi^{(n=0)}_{e}(x)$. 
The non-zero mode part   $\partial_{x}\varphi^{(n\neq0)}_{e}(x)$ is  induced by the non-uniform ground state  $-n_{e}\frac{\textbf{h(x,d)}}{2}$.  This current will be obtained within  the linear response theory \cite{Fetter}, where we have replaced the regular commutator by the Dirac   commutator  : 
\begin{eqnarray}
&& \delta<<F|\partial_{x}\hat{\varphi}_{e}(x,t))|F>>_{T}\equiv <<F|\partial_{x}\hat{\varphi}_{e}(x,t)|F>>_{T,\delta v (x,t)}- <<F|\partial_{x}\hat{\varphi}_{e}(x,t)|F>>_{T}\nonumber\\&&=\frac{i}{\hbar}\int_{-\frac{L}{2}}^{\frac{L}{2}}\,dy\int_{0}^{t}\,dt'<<F|[(-e)\delta v (y,t')(\frac{1}{\sqrt{\pi}}\partial_{y}\vartheta_{e}(y,t')  -n_{e}\frac{\textbf{h(y,d)}}{2}),\partial_{x}\hat{\varphi}_{e}(x,t)]_{Dirac}|F>>_{T}\nonumber\\&&\approx \tau\frac{e n_{e}}{\hbar\kappa}[1-\frac{\textbf{h(x,d)}}{2}] \int_{-\frac{L}{2}}^{\frac{L}{2}}\,dy \frac{\textbf{h(y,d)}}{2}\partial_{x}[\frac{1}{\sqrt{(x-y)^2+a^2 }}-\frac{1}{\sqrt{(x-y)^2+\xi^2}}]
\end{eqnarray} 
$ <<F|... |F >>_{T}$ stands for thermodynamic  expectation  value. Using the  low energy Bosonic spectrum,  which emerges from equation $(27)$,  we approximate for times $t$   obeying  $v _{eff.}\frac{2\pi}{L} t<1$, the time integration by  a constant time      $\tau$   if $v _{eff.}\frac{2\pi}{L}\tau<1$.  At  finite temperature, $\tau$ is given by  $\tau =smallest [\frac{L}{v_{eff.}},\frac{L_{T}}{v_{eff.}}]\equiv\frac{L_{T,L}}{v_{eff.}}$  
The zero mode current is computed using the current operator  $\partial_{x}\varphi^{(n=0)}_{e}(x)$. At  finite temperatures  static electric current    will be given by   the thermodynamic expectation value of the  operator   $\frac{N_{\uparrow,L}+N_{\downarrow,L}-N_{\uparrow,R}-N_{\downarrow,R}}{L}$.  This expectation value is   determined  by the thermal reservoirs. The  thermal function is represented by $<<F|(N_{\sigma,L}-N_{\sigma,R})|F>>_{T}$ ($T$ stands for the thermal occupation values). For the wire region, we have the Hamiltonian  $ H^{(n=0)}_{wire}=\frac{h v }{2L}\sum_{\sigma=\uparrow,\downarrow}(\frac{d}{L})[N^2_{\sigma,L}+N^2_{\sigma,R}]$ which is subjected to the constraint of exclusion of double occupancy. For   the leads, we  have  the $unconstrained$ Hamiltonian $H^{(n=0)}_{leads}=\frac{h v }{2L}\sum_{\sigma=\uparrow,\downarrow}(\frac{L-d}{L})[N^2_{\sigma,L}+N^2_{\sigma,R}]$. 
In the limit of long leads $\frac{d}{L}<<1$, the thermal weight of the region $d$ is negligible in comparison to the leads region.  Therefore, the  thermal expectation value   is given by the  unconstrained  Hamiltonian:
\begin{equation}
H^{(n=0)}\approx\frac{h v }{2L}\sum_{\sigma=\uparrow ,\downarrow}[N^2_{\sigma,L}+N^2_{\sigma,R}]+\sum_{\sigma=\uparrow,\downarrow}[\mu_{L}N_{\sigma,L}+\mu_{R}N_{\sigma,R}]
\label{therm}
\end{equation}
We compute the  partition function  under the assumption that $\frac{d}{L}<<1$. Therefore, the  constraint partition function  $Z=  T_{r}[ e^{- \frac{H^{(n=0)}_{wire}+H^{(n=0)}_{leads}+H^{n=0)}_{reservoir}}{K_{B}T}}]$ can be replaced by the unconstrained  partition function  $Z=  T_{r}[ e^{- \frac{H^{n=0)}}{K_{B}T}}]$  computed with the unconstrained Hamiltonian given in equation $(22)$.  
As a result, the  thermal occupation function \hspace{0.1 in}  $<<F|(N_{\sigma,L}-N_{\sigma,R})|F>>_{T}\approx\frac{1}{Z}T_{r}[ (N_{\sigma,L}-N_{\sigma,R}) e^{- \frac{H^{(n=0)}}{K_{B}T}}]$\hspace{0.1 in}   is given in terms   of  the thermal  Fermi-Dirac occupation functions   $f_{F.D.}(\epsilon_{n}-\mu_{L})=\frac{1}{1+e^\frac{(\epsilon_{n}-\mu_{L})}{K_{B}T}}$,
$f_{F.D.}(\epsilon_{n}-\mu_{R})=\frac{1}{1+e^\frac{(\epsilon_{n}-\mu_{R})}{K_{B}T}}$  expressed in terms of the  single particle spectrum $\epsilon_{n}=\frac{h v}{L}(n-\frac{1}{2})$, n=1,2,3... . \cite{Avadh,Mathieu}:
\begin{eqnarray} 
&&\sum_{\sigma=\uparrow,\downarrow}<<F|N_{\sigma,L}-N_{\sigma,R}|F>>_{T}\approx\frac{1}{Z}T_{r}[ (N_{\sigma,L}-N_{\sigma,R}) e^{- \frac{H^{(n=0)}}{K_{B}T}}]\nonumber\\&&\approx 2\sum_{n=1}^{n=\infty} [f_{F.D.}(\epsilon_{n}-\mu_{L})]-f_{F.D.}(\epsilon_{n}-\mu_{R})])\approx_{\frac{L}{L_{T}}>1}  \frac{eV }{h}(\frac{2 L}{v})
\end{eqnarray}
In order to compute the static current caused by the  static potential  $V$, we have to perform a space average which singles out the zero mode current. This can be achieved if we perform a space average over the length $L_{T,L}$.  After performing the space average, we see that on a length scale $L_{T}$  that  particle-hole current  $\hat{I}^{(n\neq0)}_{e}$  vanishes. This result is seen from the the result   $V^{(+)}_{eff.}+V^{(-)}_{eff.}=0$ given in equation $(38)$. Therefore, we find ( see equation  $(38)$) that  the thermal expectation value of the current is given by  zero mode  current  operator  $<<F|\hat{I}^{(n=0)}_{e}|F>>_{T}$ .

Using the length $L_{T,L}= smallest[L,L_{T}]$, we perform a  space  and thermodynamic  average   
$\int_{\frac{-L_{T,L}}{2}}^{\frac{L_{T,L}}{2}}\,\frac{dx}{L_{T,L}}<<F|\hat{I}_{e}|F>>_{T}$ and find:
\begin{eqnarray}
&&I_{e}=\int_{\frac{-L_{T,L}}{2}}^{\frac{L_{T,L}}{2}}\,\frac{dx}{L_{T,L}}<<F|\hat{I}_{e}|F>>_{T} \nonumber\\&& =\int_{\frac{-L_{T,L}}{2}}^{\frac{L_{T,L}}{2}}\,\frac{dx}{L_{T,L}}<<F|\hat{I}^{(n\neq0)}_{e}|F>>_{T} + \int_{\frac{-L_{T,L}}{2}}^{\frac{L_{T,L}}{2}}\,\frac{dx}{L_{T,L}}<<F|\hat{I}^{(n=0)}_{e}|F>>_{T} 
 \nonumber\\&&=(\frac{v}{v_{eff.}})(\frac{e^2  n_{e} }{h})(\frac{2\pi e}{ \kappa}) \int_{0}^{\frac{L_{T}}{2}}\,dx(1-\frac{\textbf{h(x,d)}}{2})^2\int_{-\frac{L}{2}}^{\frac{L}{2}}\,dy \frac{1}{2}\partial_{x}[\frac{ \textbf{h(y,d)}}{\sqrt{(x-y)^2+a^2 }}-\frac{\textbf{h(y,d)}}{\sqrt{(x-y)^2+\xi^2} }] 
 \nonumber\\&&+(\frac{v}{v_{eff.}})(\frac{e^2  n_{e} }{h})(\frac{2\pi e}{ \kappa}) \int_{-\frac{L_{T,L}}{2}}^{0}\,dx(1-\frac{\textbf{h(x,d)}}{2})^2\int_{-\frac{L}{2}}^{\frac{L}{2}}\,dy \frac{1}{2}\partial_{x}[\frac{ \textbf{h(y,d)}}{\sqrt{(x-y)^2+a^2 }}-\frac{\textbf{h(y,d)}}{\sqrt{(x-y)^2+\xi^2} }]
\nonumber\\&&+
\frac{ ev}{L} \int_{\frac{-L_{T,L}}{2}}^{\frac{L_{T,L}}{2}}\,\frac{dx}{L_{T,L}}[1-\frac{ \textbf{h}(x,d)}{2}]\sum_{\sigma=\uparrow,\downarrow}<<F|N_{\sigma,L}-N_{\sigma,R}|F>>_{T}\nonumber\\&&= 2\frac{e^2 }{h}(V^{(+)}_{eff.}+V^{(-)}_{eff.})+[1-\int_{\frac{-L_{T,L}}{2}}^{\frac{L_{T,L}}{2}}\,\frac{dx}{L_{T,L}}\frac{ \textbf{h}(x,d)}{2}](2\frac{e^2}{h}V)=[1-\int_{\frac{-L_{T,L}}{2}}^{\frac{L_{T,L}}{2}}\,\frac{dx}{L_{T,L}}\frac{ \textbf{h}(x,d)}{2}](2\frac{e^2}{h}V)\nonumber\\&&=(1-0.5\frac{d}{L_{T,L}})(2\frac{e^2}{h})V
\end{eqnarray}
$ V^{+}_{eff.}$  and   $ V^{-}_{eff.}$ represent  the opposite voltage drop on each side  of the wire    which, due to the symmetry  $\textbf{h(x,d)}= \textbf{h(-x,d)}$ obeys,   $V^{+}_{eff.}+ V^{-}_{eff.}=0$.
\begin{eqnarray}
&&V^{(+)}_{eff.} =-V^{(-)}_{eff.}=n_{e}(\frac{v}{v_{eff.}})(\frac{\pi e}{ \kappa})\int_{0}^{\frac{L_{T,L}}{2}}\,dx(1-\frac{\textbf{h(x,d)}}{2})^2[\int_{-\frac{L}{2}}^{\frac{L}{2}}\,dy \frac{1}{2}\partial_{x}[\frac{ \textbf{h(y,d)}}{\sqrt{(x-y)^2+a^2 }}-\frac{\textbf{h(y,d)}}{\sqrt{(x-y)^2+\xi^2} }]\nonumber\\&&\equiv n_{e}(\frac{v}{v_{eff.}})(\frac{\pi e}{ \kappa})W[\frac{L}{d},\frac{\xi}{a}]
\end{eqnarray}
In the present case, due to the symmetry, we can not observe the  voltage $n_{e}(\frac{v}{v_{eff.}})(\frac{\pi e}{ \kappa})W[\frac{L}{d},\frac{\xi}{a}]$.   We believe that  for   an asymmetric situation this  voltage  might be measured.  For the  case where  half of the wire-leads  system  $x\geq\frac{d}{2}$ is controlled by the strong interaction $U=\infty$ and the other half $x<\frac{d}{2}$ is  non-interacting    $U=0$, the conductance  will be affected  by  the voltage $ V^{+}_{eff.}$. This voltage is in the range of $10^{-9} volts $ and might be measurable. In figure $3$ we show  the   potential $W[\frac{L}{d},\frac{\xi}{a}]$ as a function of the screening length.

Equation $(38)$ shows that the conductance for $\frac{d}{L_{T,L}}=1$  is given by  $G=2( 0.5 \frac{e^2}{h})$, reflecting the fact that the anomalous commutator  for the wire has been modified to $\frac{i}{2}\hbar$.
For different lengths the conductance is modified. Following \cite{Siljuasen}, we  investigate  the  dependence of the conductance in terms of $\frac{L}{D}$ and  $\frac{d}{L_{T}}$ (the effect   of  the overlapping region  $\epsilon$  between the leads and the wire studied by  \cite{Siljuasen} will be neglected here).
The thermal length $L_{T}\equiv\frac{\hbar v_{eff.}}{K_{B}T}=\frac{v_{eff.}}{v} \frac{\hbar v}{K_{B}T}=(1-\frac{3d}{4L})L^{(0)}_{T}$ where \textbf{ $L^{(0)}_{T}$ is the thermal length defined in the leads}. 
The length $L_{T,L}\approx L_{T}\leq L$ is given  by the crossover function:  $\frac{1}{L_{T,L}}\approx \frac{1}{L_{T}}[1+\frac{L_{T}}{L}]$. As  a result we find for   the conductance:

\begin{eqnarray}
&&G[\frac{d}{L},\frac{T_{d}}{T};U=\infty]=(2\frac{e^2}{h})[1-\frac{1}{2}\frac{ T}{T_{d}}\frac{1}{(1-\frac{3}{4}\frac{d}{L})}(1+(\frac{T_{d}}{T})(\frac{d}{L})(1-\frac{3}{4}\frac{d}{L}))]\nonumber\\&&
\end{eqnarray}
The result, of equation $(40)$ is shown in figure $2$. We observe  that when the   length of the wire  $d=\frac{\hbar v}{K_{B}T_{d}}$  is in the range  $\frac{T_{d}}{T}\approx 1-2$ the conductance is anomalous in agreement with the experimental results reported in ref. \cite{Reilly}.  (In the experiments the anomalous conductance is obtained as a result of the variation of the gate voltage,  we will interpret  this result as the region where the strong coupling point  $U=\infty$ has been reached. At this point, when the temperature or the length are  varied the conductance varies. We believe that this variation  of the  conductance is demonstrated in figure $2$.)

\vspace{0.2 in}

\textbf{7. The effect of the Zeeman magnetic field}

\vspace{0.2 in} 

A complete discussion of the anomalous conductance must include the ferromagnetic interaction proposed by  \cite{Han}. According to \cite{Han}, the emergence of the $ 0.7$ conductance anomaly is due to ferromagnetic interactions.
Since the effect of constraints takes into  consideration the exclusion of double occupancy, which also holds for the ferromagnetic case, we believe that our results include, in a qualitative  way, the results obtained in ref. \cite{Han}.

The results reported by \cite{Han} show that in the presence of Zeeman magnetic field the 
$0.7$ plateau evolves into a robust $0.5$ plateau.
In our case, the value of the conductance varies from  $0.5$ to $0.7$ (in the absence of the Zeeman interaction). These values are  determined by the  function  $\textbf{h}(x,d)$  the temperature and length of the wire. 

In order to consider the effect of a Zeeman interaction, we observe that the Fermi velocity $v$ is replaced by: $v^{\uparrow}=v+\frac{\Delta}{2}$,  $v^{\downarrow}=v-\frac{\Delta}{2}$ where $\Delta= (\frac{\hbar}{ m_{e}})\overline{K}_{F}[\sqrt{1+\frac{g_{||}\mu_{B}B_{||}}{E_{F}}}-\sqrt{1-\frac{g_{||}\mu_{B}B_{||}}{E_{F}}}]$ represents the Zeeman interaction, $\overline{K}_{F}=\frac{K^{\uparrow}_{F}+K^{\downarrow}_{F}}{2}$ is the Fermi momentum, $B_{||}$  is the magnetic field and $m_{e}$ is the electronic mass.     
As a result, the Hamiltonian  for the wire takes the form:
\begin{equation}
H_{wire}=\int_{\frac{-d}{2}}^{\frac{d}{2}}\,dx([\frac{\hbar (v+\frac{\Delta}{2})}{2}(\partial_{x}\varphi _{\uparrow}(x))^2+(\partial_{x}\vartheta_{\uparrow}(x))^2]+ [\frac{\hbar (v-\frac{\Delta}{2})}{2}(\partial_{x}\varphi _{\downarrow}(x))^2+(\partial_{x}\vartheta_{\downarrow}(x))^2])
\label{Zeeman}
\end{equation}
This Hamiltonian must be accompanied by the constraint operator of exclusion of double occupancy.
\begin{equation}   
\Psi_{\uparrow}(x) \Psi_{\downarrow}(x)|F>= 2e^{-i\sqrt{4\pi}\varphi_{e}(x)} Q^{Zeeman}_{1}(x)|F>=0; \hspace{0.1 in}  \frac{-d}{2}\leq x\leq\frac{d}{2} 
\label{contz}
\end{equation}  
 where   $ Q^{Zeeman}_{1}(x)$ is  the Bosonic   constraint in the presence of the Zeeman interaction:  
\begin{eqnarray}
Q^{Zeeman}_{1}(x)&=&cos[2\overline{K}_{F}x +\sqrt{2\pi}\vartheta_{e}(x)]+cos[(\frac{m_{e}}{\hbar})\Delta x+\sqrt{2\pi}\vartheta_{s}(x)];\hspace{0.2 in} Q^{Zeeman}_{1}(x)|F>=0\nonumber\\&&
\label{zoperator}
\end{eqnarray}
The Hamiltonian and the constraint operator show that neither the spin current nor the charge  current are conserved. For a finite repulsive interaction  and next nearest neighbor, this problem has been studied  in the past \cite{hsuan}. One finds two different  Luttinger liquids for spin up and spin down, which give rise to two different conductances as  functions of the spin polarization \cite{hsuan} (see equations $(26-27)$ and figures $(1-2)$ in reference \cite{hsuan}). 

For the present model $U=\infty$, we observe the following:  when  the Zeeman interaction $\Delta$ is small, we do not expect  significant changes  with respect  to the results obtained for  $\Delta=0$.

For large values of  the  Zeeman interaction   , $\frac{g_{||}\mu_{B}B_{||}}{E_{F}}>1$,    the effect of the constraint is negligible. This can be seen  from the  bias field  $\Delta x$ which induces a space dependent oscillation for   the constraint  $ Q^{Zeeman}_{1}(x)$   in equation $(43)$ . Due to the  oscillations  we obtain, $Q^{Zeeman}_{1}(x)\approx 0$. 
The wire Hamiltonian is replaced by  an unconstrained  polarized   wire  
$H_{wire}\approx H^{(\uparrow)}_{wire}=\int_{\frac{-d}{2}}^{\frac{d}{2}}\,dx[\frac{\hbar (v+\frac{\Delta}{2})}{2}(\partial_{x}\varphi _{\uparrow}(x))^2+(\partial_{x}\vartheta_{\uparrow}(x))^2]$\hspace{4.0 in}
Therefore, we  recover   the  robust $0.5$ plateau in agreement with \cite{Han}. 

\vspace{0.2 in}
 
\textbf{8.  Conclusion}

\vspace{0.2 in} 

We have solved  the  problem   of exclusion of double occupancy using  Dirac's method for constraints. We have found that the  anomalous commutation rules are modified causing the conductance to be anomalous.  Applying  this theory to quantum wires, we show that our theory can explain the anomalous  conductance observed  by  \cite{Pepper,Cronennwett,Michael,Picciotto}  and is in agreement with the Monte-Carlo simulation  reported in refs.\cite{Siljuasen,Han}.

\clearpage
\begin{figure}
\begin{center}
\includegraphics[width=7.5 in ]{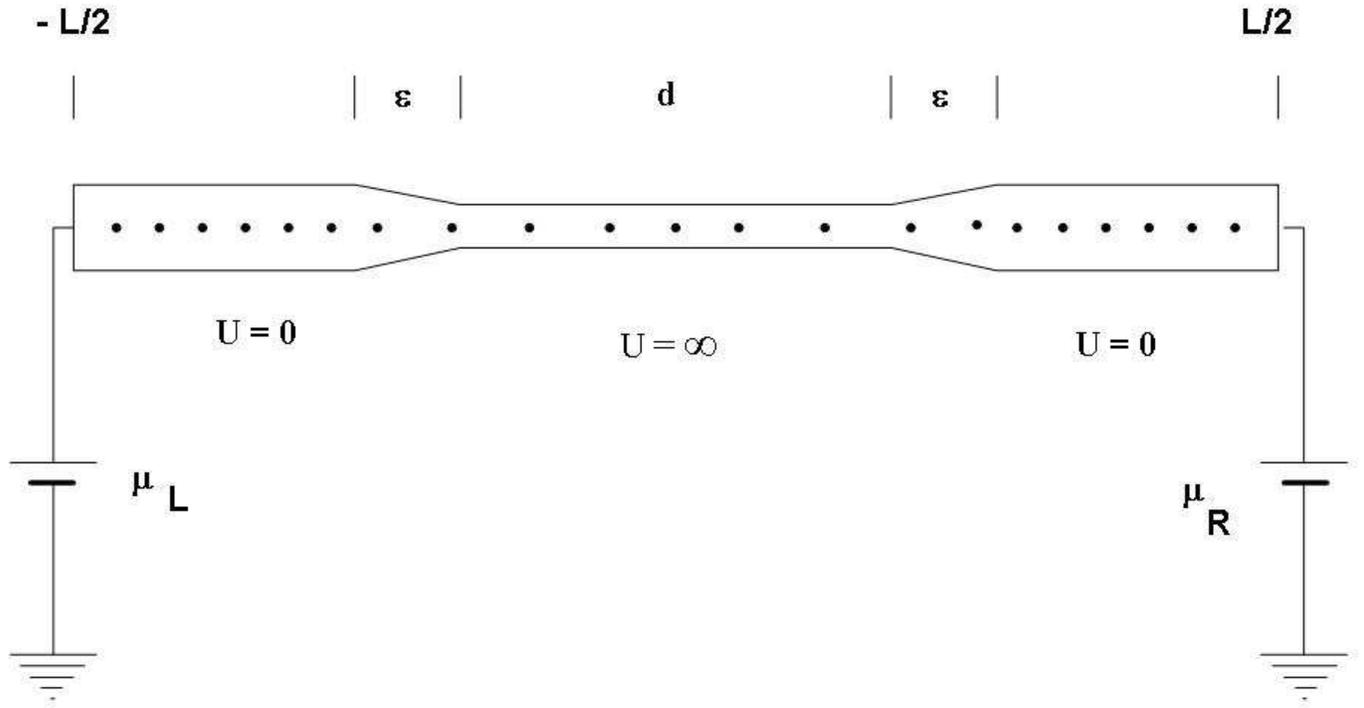}
\end{center}
\caption{The 1d conducting channel:  a) $noninteracting$  leads $U=0$    regions $|x|>\frac{d}{2}$ , $strongly$ $interacting$ region  $U=\infty$ restricted  to $ |x|\leq \frac{d}{2}$ and contact region $\epsilon$.} 
\end{figure}
\clearpage

\begin{figure}
\begin{center}
\includegraphics[width=7.0 in ]{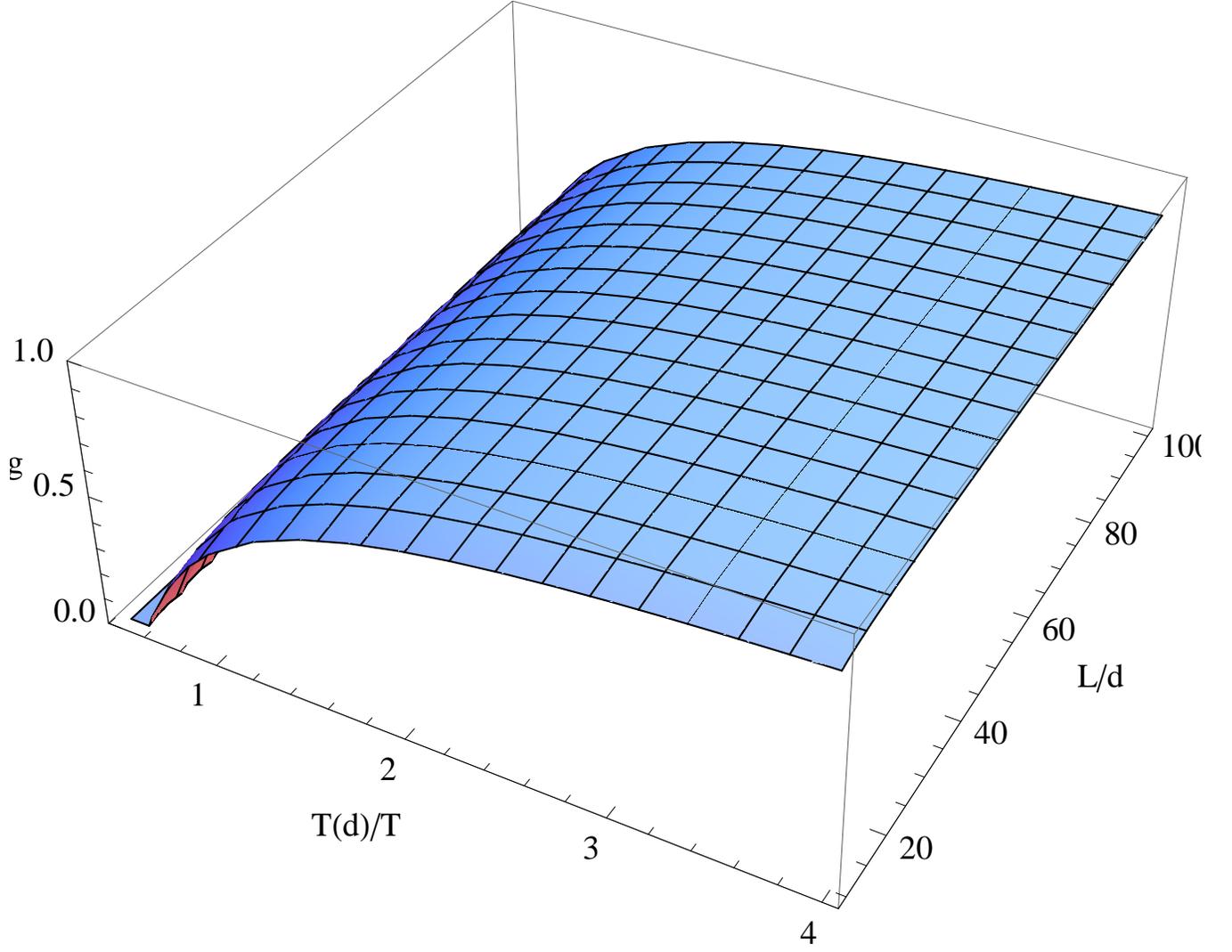}
\end{center}
\caption{The Conductance $g(U=\infty)\equiv\frac{G[\frac{T_{d}}{T},\frac{L}{d};U=\infty]}{{2 e^2 }{h}}$  as a function of $\frac{T_{d}}{T}$ and $\frac{L}{d}$  where $T_{d}=\frac{\hbar v}{d K_{B}}$.}   
\end{figure}

\clearpage

\begin{figure}
\begin{center}
\includegraphics[width=7.0 in ]{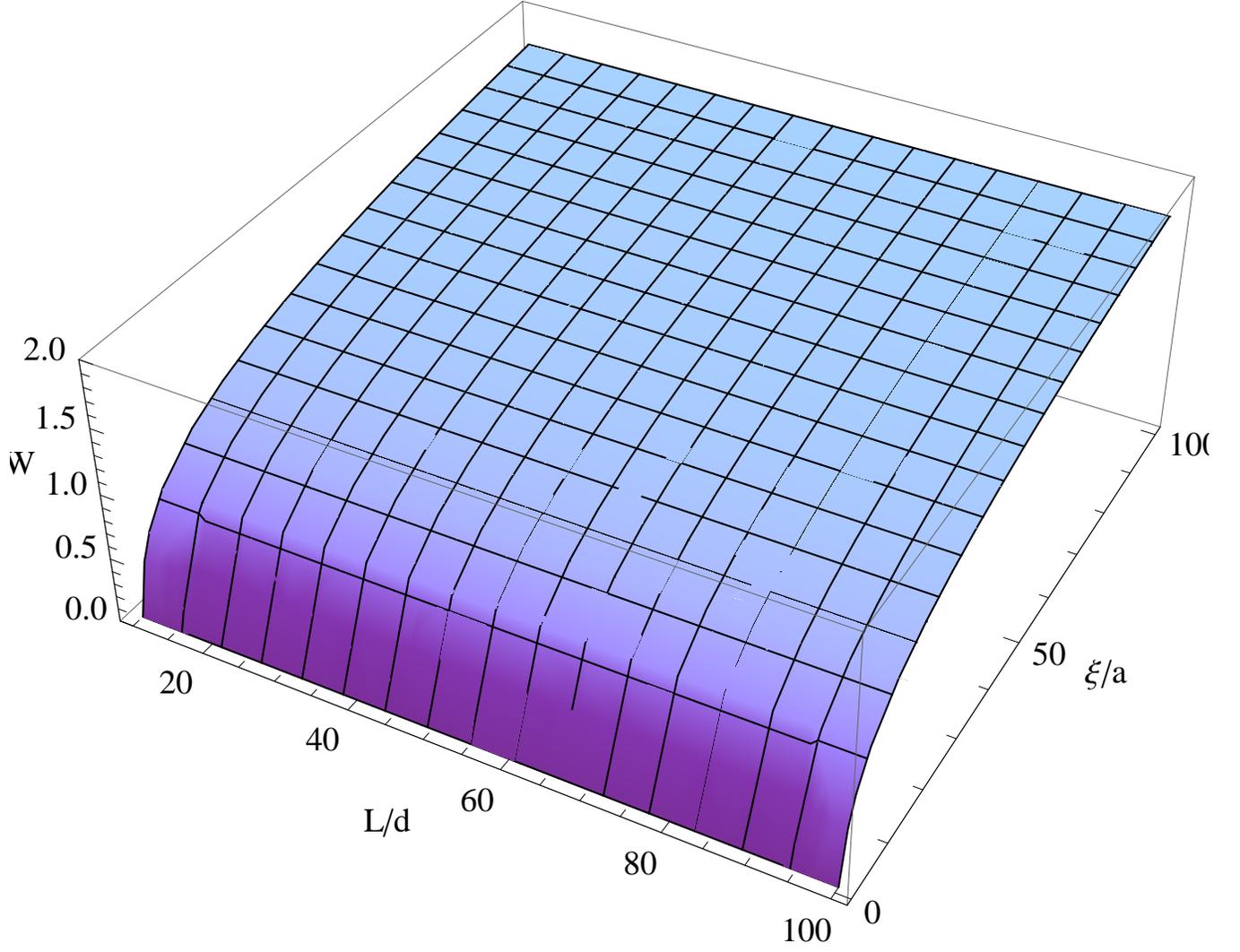}
\end{center}
\caption{The effective voltage on each side of the wire $V^{(+)}\equiv  n_{e}(\frac{v}{v_{eff.}})(\frac{\pi e}{ \kappa})W[\frac{L}{d},\frac{\xi}{a}]$  where $W[\frac{L}{d},\frac{\xi}{a}]$ is a dimensionless function.}   
\end{figure}

\end{document}